\documentclass[12pt]{article}

\usepackage{epsfig}

\def\beq{\begin{equation}}
\def\eeq{\end{equation}}
\def\bea{\begin{eqnarray}}
\def\eea{\end{eqnarray}}

\def\tev{\, {\rm TeV}}
\def\gev{\, {\rm GeV}}

\newcommand{\gsim}{\lower.7ex\hbox{$\;\stackrel{\textstyle>}{\sim}\;$}}
\newcommand{\lsim}{\lower.7ex\hbox{$\;\stackrel{\textstyle<}{\sim}\;$}}
\def\tenrep{{\bf 10}}
\def\fiverep{{\bf 5}}
\def\fivebarrep{\overline{\bf 5}}
\def\twentyfourrep{{\bf 24}}
\def\singletrep{{\bf 1}}
\def\mpl{M_{\rm Pl}}
\def\gauge{{\cal W}}
\def\mbar{{\overline{M}}}

\begin{document}

\setlength{\baselineskip}{0.25in}


\begin{titlepage}
\noindent
\begin{flushright}
MCTP-03-56 \\
\end{flushright}
\vspace{1cm}

\begin{center}
  \begin{Large}
    \begin{bf}
Gravity-assisted exact unification in minimal \\
supersymmetric $SU(5)$ and its gaugino mass spectrum

    \end{bf}
  \end{Large}
\end{center}
\vspace{0.2cm}
\begin{center}
\begin{large}
Kazuhiro Tobe$^{a,b}$, James D. Wells$^{a}$ \\
\end{large}
  \vspace{0.3cm}
  \begin{it}
${}^{(a)}$Michigan Center for Theoretical Physics (MCTP) \\
        ~~University of Michigan, Ann Arbor, MI 48109-1120, USA \\
\vspace{0.1cm}
${}^{(b)}$Department of Physics, University of California, Davis 95616, USA 
  \end{it}

\end{center}

\begin{abstract}

Minimal supersymmetric $SU(5)$ with exact unification is naively
inconsistent with proton decay constraints. However, it can be made
viable by a gravity-induced non-renormalizable operator connecting
the adjoint Higgs boson and adjoint vector boson representations.
We compute the allowed coupling space for this theory and find
natural compatibility with proton decay constraints even for relatively
light superpartner masses.  The modifications away from the naive
$SU(5)$ theory have an impact on the gaugino mass spectrum, which we calculate.
A combination of  precision
Linear Collider and Large Hadron Collider 
measurements of superpartner masses would enable
interesting tests of the high-scale form of minimal 
supersymmetric $SU(5)$.

\end{abstract}

\vspace{1cm}

\begin{flushleft}
hep-ph/0312159 \\
December 2003
\end{flushleft}

\end{titlepage}



The three gauge couplings of the minimal supersymmetric standard model (MSSM)
unify to within 1\% of each other at a high scale 
$\sim 2\times 10^{16}\gev$.  Simple Grand Unified Theories (GUTs) predict
such an outcome, where the low-scale gauge couplings must flow to within
a small neighborhood of each other (less than few percent mismatch)
at the high scale.
Exact unification occurs only when all threshold corrections
at the high scale are properly taken into account.

The simplest supersymmetric GUT model is minimal $SU(5)$, with 
matter representations $\{ \tenrep_i,\fivebarrep_i, \singletrep_i\}$,
the gauge boson representation $\twentyfourrep$,
and Higgs representations $\{ \twentyfourrep_H,\fiverep_H,\fivebarrep_H\}$.
Precise gauge coupling unification at the high-scale must take into
account threshold corrections from heavy GUT remnants of the $\twentyfourrep$,
$\twentyfourrep_H$, and $\fiverep_H+\fivebarrep_H$ representations.
The colored Higgsino triplets $H_c$ from the $\fiverep_H+\fivebarrep_H$ 
representations also contribute to dangerous dimension five operators
mediating proton decay~\cite{Sakai:1981pk}. A careful analysis of both
gauge coupling 
unification and proton decay in minimal supersymmetric $SU(5)$
concludes
\bea
M_{H_c} & \gsim & 10^{17}\gev~~~({\rm from~proton~decay~constraints}) 
\label{MHC_proton_limit}
\\
M_{H_c} & \simeq & {\rm few}\,\times 10^{15}\gev~~~
({\rm from~gauge~coupling~unification~constraints})
\label{MHC_gcc_limit}
\eea
if superpartner masses are in the TeV region.  This has led to the
conclusion that minimal $SU(5)$ is
dead~\cite{Goto:1998qg,Murayama:2001ur,Raby:2002wc}
or perhaps at least highly constrained with superpartner masses in the
$10\tev$ range~\cite{Bajc:2002bv} which strains its ability
to naturally explain the electroweak symmetry breaking scale.
\footnote{There exist other $SU(5)$ models that are consistent
with both gauge coupling unification and proton decay. For example,
see Ref.~\cite{Hisano:1994fn}.}

In this letter we wish to point out two conclusions we have come
to recently, apropos to the
discussion above.  First, similar to the effects
found in Refs.~\cite{Hill:1983xh}-\cite{Huitu:1999eh} we have found that
expected non-renormalizable gauge-kinetic operators
in the GUT theory can redeem minimal $SU(5)$ without requiring
unnaturally large coefficients.  Second, we have computed the imprint
of this effect on the gaugino masses, and found 
the resulting 
magnitudes of their relative shifts at the GUT scale to be within
the sensitivities of future and planned colliders.

Minimal
$SU(5)$ as a purely renormalizable supersymmetric theory was never
viable because
unification of down-quark Yukawa couplings with lepton Yukawa couplings
does not work for the first two generations.  It has been understood
for a very long time now that non-trivial
non-renormalizable operators
(NROs) are needed. This is no extraordinary
burden on the theory, however, as the Planck scale is not far from the GUT
scale and NROs
induced by supergravity are expected
and of sufficient size to implement the flavor gymnastics
required to reproduce the masses and mixings of the quarks and leptons.

It has also been known for some time that NROs
can dramatically affect gauge coupling unification and gaugino
masses~\cite{Hill:1983xh}-\cite{Huitu:1999eh}.  
These operators should not necessarily
be viewed as sources of GUT-scale obfuscation, but rather as
potential saviors for a theory that struggles to survive without
them.   Minimal $SU(5)$ is one such theory.

We can write the gauge-kinetic function of minimal $SU(5)$ as
\beq
\int d^2\theta\left[ \frac{S}{8\mpl}\gauge\gauge +
\frac{y \Sigma}{\mpl}\gauge\gauge\right]
\label{gkf}
\eeq
where $\Sigma=\twentyfourrep_H$ and
$\langle S\rangle = \mpl/g_G^2+\theta^2 F_S$ contains the
effective singlet supersymmetry breaking.  The $SU(5)$ gauge
coupling is $g_G$ and the universal contribution to the masses 
of all gauginos is
$M_{1/2}=-g_G^2F_S/(2\mpl)$. 

This second term of Eq.~(\ref{gkf}) is the focus of our
analysis\footnote{See Refs.~\cite{Bajc:2002bv, Emmanuel-Costa:2003pu} 
for discussion of other types of NROs useful to cure the minimal
$SU(5)$ problem.} as it connects the adjoint Higgs representation
to the gauge fields via a NRO.  Not only is the operator expected,
but it is guaranteed to contribute to the gauge coupling corrections
because the adjoint Higgs must get a vacuum expectation value (vev) 
of the form 
\beq
\langle \Sigma\rangle = v_\Sigma\, {\rm diag}\, 
\left( \frac{2}{3},\frac{2}{3},\frac{2}{3},-1,-1\right)
\eeq
to break $SU(5)$ to $SU(3)\times SU(2)_L\times U(1)_Y$ at
the GUT scale.  The numerical value of $v_\Sigma$ depends on details
of the couplings but should be around the GUT scale of $10^{16}\gev$.  

The relationships between the GUT scale gauge coupling $g_G$ and
the low-scale gauge couplings $g_i(Q)$ of the
MSSM effective theory are
\beq
\frac{1}{g^2_i(Q)}=\frac{1}{g_G^2(Q)}+\Delta^G_i(Q)+c_i\epsilon
\label{gcc}
\eeq
where $\epsilon=8y v_\Sigma/\mpl$ and $c_i=\{ -1/3,-1,2/3\}$ for the gauge
groups $i=\{ U(1)_Y, SU(2)_L,SU(3)\}$ respectively. Here we adopt the
GUT normalized $U(1)_Y$ gauge coupling $g_1^2=(5/3)g_Y^2$.
The $\Delta^G_i(Q)$ functions are the threshold corrections due to
heavy GUT states; $\Delta^G_i(Q)=1/(8\pi^2) \sum_a b_{a i} \ln (Q/M_a)$
where $b_{a i}$ and $M_a$ are $\beta$ function coefficient of a heavy
particle and its mass, respectively. They are explicitly written by
\bea
\Delta^G_1(Q) & = & \frac{1}{8\pi^2}\left( -10\ln \frac{Q}{M_V}
+\frac{2}{5}\ln\frac{Q}{M_{H_c}}\right) \\
\Delta^G_2(Q) & = & \frac{1}{8\pi^2}\left( -6\ln \frac{Q}{M_V}
+2\ln\frac{Q}{M_{\Sigma}}\right) \\
\Delta^G_3(Q) & = & \frac{1}{8\pi^2}\left( -4\ln \frac{Q}{M_V}
+\ln\frac{Q}{M_{H_c}}+3\ln\frac{Q}{M_{\Sigma}}\right) .
\eea
We will be working with this equation near the GUT scale, $Q\sim 10^{16}\gev$,
and so the couplings $g_i(Q)$ are assumed to be those that have
been measured at the weak scale, renormalized by weak-scale supersymmetric
particle threshold corrections and run up to the high scale $Q$ using
two-loop renormalization group evolution~\cite{Bagger:1995bw}.
Our equations are always in the $\overline{\rm DR}$ scheme.

It is easy to
see how the triplet Higgsino mass is severely constrained by
unification requirements. Let's consider the $\epsilon=0$ case for a
moment. There exists a linear combination of
$g_i^{-2}$ that depends only on $M_{H_c}$ 
and not on the other unknown GUT scale states~\cite{Hisano:1992mh}:
\beq
-\frac{1}{g^2_1(Q)}+\frac{3}{g^2_2(Q)}-\frac{2}{g^2_3(Q)}
= \frac{3}{5\pi^2}\ln \frac{M_{H_c}}{Q}.
\label{gi combo}
\eeq
This equation is true for any scale $Q$ at the one-loop level, but it is
most instructive to evaluate it at the unification scale $\Lambda_U$,
which we define to be the place where
$g_1(\Lambda_U)=g_2(\Lambda_U)=g_U$, 
\beq
\frac{1}{g^2_U}-\frac{1}{g^2_3(\Lambda_U)}
= \frac{3}{10\pi^2}\ln \frac{M_{H_c}}{\Lambda_U}.
\label{gu combo}
\eeq
$\Lambda_U$ depends mildly on the low-scale superpartner
masses, but it is always within the range
\beq
1\times 10^{16}\gev\lsim \Lambda_U\lsim 2\times 10^{16}\gev
\eeq
for superpartner masses at the TeV scale and below.

It is well
known~\cite{Bagger:1995bw} 
that $g_3(\Lambda_U)<g_U$, albeit by less than 1\%. Nevertheless,
this implies that the LHS of Eq.~(\ref{gu combo}) is necessarily negative.
We see that $M_{H_c}<\Lambda_U$ is required for the RHS to be negative
and successful gauge coupling
unification to occur.  But this is in conflict with the proton decay
requirement that $M_{H_c}>10^{17}\gev (> \Lambda_U)$.

However, non-zero $\epsilon~(>0)$ can easily and naturally
enable a large $M_{H_c}$. 
Because of an interesting and non-trivial relation between
$c_i$ and a $\beta$-function coefficients $b_{H_ci}=\{2/5,0,1\}$ of $H_c$
($c_i=-{\bf 1}_i+\frac{5}{3} b_{H_ci}$), an inclusion of non-zero
$\epsilon$ only affects the unified gauge coupling and color-triplet
Higgsino mass $M_{H_c}$ as can be seen from Eq.~(\ref{gcc}). In other
words, if we define
the effective colored Higgsino mass to be
$M^{\rm eff}_{H_c}=M_{H_c}\exp{(-40\pi^2 \epsilon/3)}$, the above 
constraints discussion
in the case with $\epsilon=0$ applies to $M^{\rm eff}_{H_c}$.  
Therefore, even though the effective
colored Higgsino mass $M^{\rm eff}_{H_c}$ is severely constrained by
gauge coupling unification (Eq.~(\ref{MHC_gcc_limit})), the real
colored Higgsino mass $M_{H_c}=M^{\rm eff}_{H_c}\exp{(40\pi^2 
\epsilon/3)}$ can be large enough to satisfy the proton decay limit
Eq.~(\ref{MHC_proton_limit}) if $\epsilon$ is positive and 
of order a few percent.\footnote{
Other heavy particle masses are constrained by the gauge coupling
unification as $9\times 10^{15}$ GeV $< (M_\Sigma M_V^2)^{1/3} <
2\times 10^{16}$ GeV. However, this constraint does not change
even if non-zero $\epsilon$ is taken into account because of
the relation $c_i=-{\bf 1}_i+\frac{5}{3} b_{H_ci}$.}
We remark also that the unified coupling governing dimension six
proton decay operators is reduced by $g^2_{G,\epsilon}/g^2_{G,0}\simeq
1-\epsilon/2$, thus increasing the proton lifetime.

\begin{figure}[t]
\centering
\includegraphics*[width=13cm]{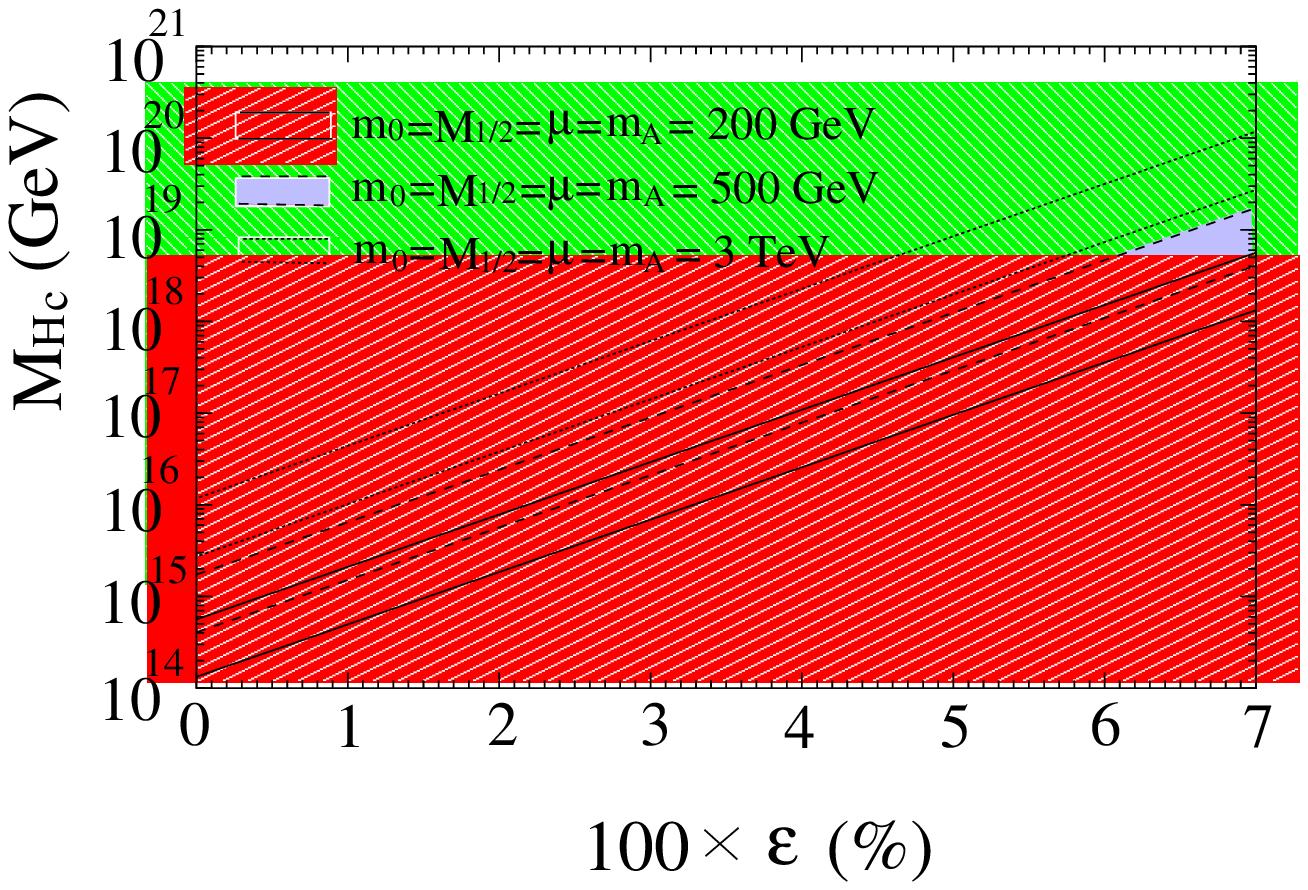}
\caption{Fit for the heavy triplet Higgsino mass as a function of adjoint-Higgs
corrections ($\epsilon$) in order to accomplish gauge coupling unification.
Here we define $m_0$ (universal scalar mass) and $M_{1/2}$ 
(universal gaugino mass) at the GUT scale, and $\mu$ and $m_A$
at the weak scale without imposing a radiative electro-weak symmetry
breaking condition.
One expects $\epsilon\sim {\rm few}\% $, and thus $M_{H_c}>10^{17}\gev$
can be naturally achieved as is required by proton decay constraints.
The width of each band is primarily due to the current uncertainty
in $\alpha_s(m_Z)$.}
\label{MC}
\end{figure}

We have done the precise numerical work to test this supposition
and the results are presented in Fig.~\ref{MC}, where the
relationship between $\epsilon$ and $M_{H_c}$ for exact unification
is established.  Each band is for a given assumed superpartner spectrum,
and the width of the band is primarily due to the current uncertainty in
$\alpha_s(m_Z)$ which we take to be $0.115<\alpha_s(m_Z)<0.119$. 

We see from the numerical
results (Fig.~\ref{MC}) that if we ignore the adjoint-Higgs NRO
correction ($\epsilon=0$) 
the triplet Higgsino mass needed for unification
is less than about $10^{16}\gev$, even for all
superpartner masses up to $3\tev$.  However, if $\epsilon\simeq {\rm few}\,\%$
we find that $M_{H_c}$ can be comfortably greater than $10^{17}\gev$, thus
enabling precision gauge coupling unification and a sufficiently 
long-lived proton.  This successful region of parameter space requires
$v_\Sigma/\mpl\simeq {\rm few}\,\%$, which is consistent
with the expectation $v_\Sigma\simeq \Lambda_U$.
It should be stressed that this is a built-in mechanism to
increase $M_{H_c}$ naturally in minimal $SU(5)$ model, and more generally,
in $SU(5)$ models in which the ${\bf 24}_H$ breaks $SU(5)$.

At present there is no known way to experimentally verify
minimal $SU(5)$, or any other GUT for that matter.  However, it
is possible to test the theory nontrivially.  On the surface it
may appear unlikely that any shifting around of $\epsilon$ and $M_{H_c}$ at
the high-scale to obtain compatibility with low-scale gauge coupling
measurements would have any discernible experimental 
implications. However, precision gaugino mass measurements do provide
a interesting probe of the framework.

One crucial realization is that the $\twentyfourrep_H$ representation vev is 
not just a scalar vev, but a vev in superspace when we take into
account the entire chiral superfield,
\beq
\langle \hat\Sigma\rangle \simeq (v_\Sigma+F_\Sigma\theta^2)\, {\rm diag}\, 
\left( \frac{2}{3},\frac{2}{3},\frac{2}{3},-1,-1\right)
\eeq
Just as the $\hat H_u$ and $\hat H_d$ Higgs superfields 
of the MSSM pick up auxiliary
field vevs when their scalar components condense, 
the $\hat \Sigma$ superfield obtains a superspace vev.

The superpotential and soft lagrangian terms we assume are
\beq
W = \frac{1}{2}M_\Sigma {\rm Tr}\, \Sigma^2 +\frac{f}{3}{\rm Tr}\, 
\Sigma^3 + M_5\fiverep_H\fivebarrep_H +\lambda \fivebarrep_H \Sigma
\fiverep_H  +\cdots
\eeq
\beq
-{\cal L}_{\rm soft} = \frac{1}{2}B_\Sigma M_\Sigma {\rm Tr}\,
\Sigma^2 + \frac{f}{3}A_\Sigma{\rm Tr}\, \Sigma^3 
+B_5M_5\fiverep_H\fivebarrep_H 
+A_\lambda \lambda \fivebarrep_H \Sigma \fiverep_H + h.c. +\cdots
\eeq
where upon minimizing the full potential we find
\bea
F_\Sigma \simeq v_\Sigma (A_\Sigma-B_\Sigma) =\frac{\epsilon \mpl}
{8y} (A_\Sigma-B_\Sigma)
\eea
which generates a correction to gaugino masses 
via the NRO in Eq.~(\ref{gkf}).

\begin{figure}[t]
\centering
\includegraphics*[width=13cm]{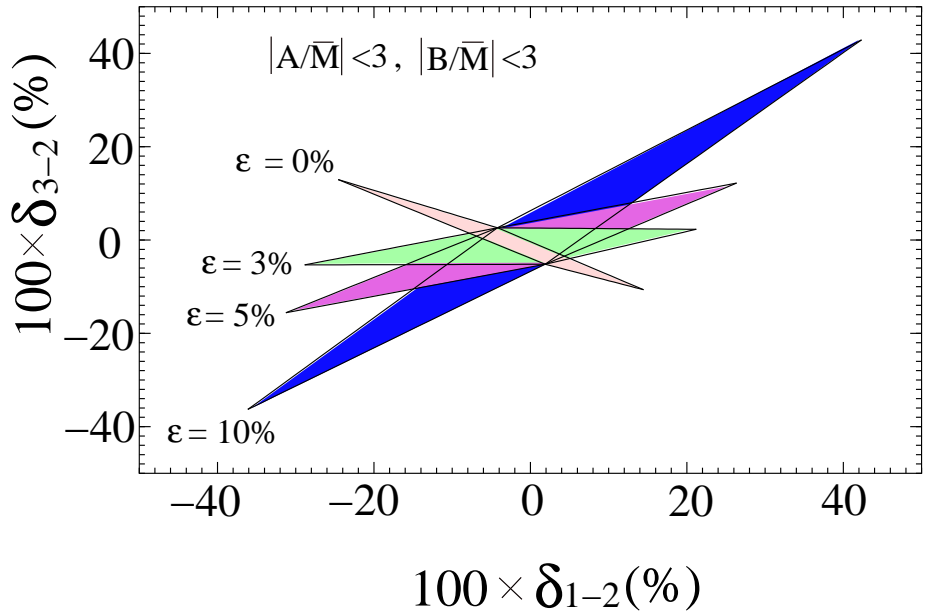}
\caption{$\delta$ corrections to the gaugino masses at the scale $\Lambda_U$
where $g_1(\Lambda_U)=g_2(\Lambda_U)$. We have defined
$\delta_{1-2}=(M_1(\Lambda_U)-M_2(\Lambda_U))/M_2(\Lambda_U)$
and $\delta_{3-2}=(M_3(\Lambda_U)-M_2(\Lambda_U))/M_2(\Lambda_U)$.
Here we have assumed universal $A$-terms $(A_\Sigma=A_\lambda\equiv A)$
and $B$-terms $(B_\Sigma =B_5\equiv B)$ and varied them over the ranges 
$|A/\overline{M}|<3$ and $|B/\overline{M}|<3$.}
\label{d12_d32}
\end{figure}

This non-zero $F_\Sigma$-term vev contributes non-universally
to each of the gaugino masses.  Taking these shifts into account
and the GUT scale threshold corrections\footnote{
We found some discrepancies in Eq.~(10) in Ref.~\cite{Hisano:1993zu}.
One is an overall sign of the second parenthetic term in the right-hand side of
Eq.~(10) which comes from the finite corrections of heavy GUT particles. 
Subsequent equations indicate that this is merely a typo.
The other is the $\delta m$ term in their Eq.~(10),
which we believe should be $-\delta m/2$.
This discrepancy originates from their Eq.~(7), where we believe the $\delta m$
in the matrix should also be $-\delta m/2$.}~\cite{Hisano:1993zu} on 
gaugino masses,
we find that the values of the gaugino masses at $\Lambda_U$ are
\bea
M_1(\Lambda_U) & = & g_U^2\overline{M}+g_U^2 \left[
\frac{1}{6}\epsilon (A_\Sigma-B_\Sigma)
-\frac{1}{16\pi^2}\left( 10g_U^2\overline{M}+10\{A_\Sigma
-B_\Sigma\} +\frac{2}{5}B_5\right)\right] \\
M_2(\Lambda_U) & = & g^2_U\overline{M}+g_U^2 \left[
\frac{1}{2}\epsilon (A_\Sigma-B_\Sigma)
-\frac{1}{16\pi^2}\left( 6g_U^2\overline{M}+6A_\Sigma
-4B_\Sigma \right)\right] \\
M_3(\Lambda_U) & = & g^2_3(\Lambda_U)\overline{M}+g_U^2 \left[
-\frac{1}{3}\epsilon (A_\Sigma-B_\Sigma)
-\frac{1}{16\pi^2}\left( 4g_U^2\overline{M}+4A_\Sigma
-B_\Sigma+B_5 \right)\right] 
\eea
where $\overline{M}=-F_S/(2 \mpl)\sim {\cal O}(m_z)$ is the supersymmetry
mass scale from the singlet field $F$-term in Eq.~(\ref{gkf}). For
our subsequent numerical work that will culminate in Fig.~\ref{d12_d32}
we use $g_U=0.711$ and $g_3(\Lambda_U)=0.705$.  These numerical values change
slightly with the superpartner masses, but the qualitative features
of the results stay the same. Furthermore, as we shall emphasize,
these quantities are unambiguously
calculable given knowledge of the low-energy superpartner spectrum.

The overall scale of the gaugino masses cannot be predicted; however,
there are some interesting correlations among ratios of the gauginos.
It is convenient to define the quantities
\beq
\delta_{1-2}=\frac{M_1(\Lambda_U)-M_2(\Lambda_U)}{M_2(\Lambda_U)}~~~{\rm and}~~~
\delta_{3-2}=\frac{M_3(\Lambda_U)-M_2(\Lambda_U)}{M_2(\Lambda_U)}.
\eeq
The $\delta$'s are defined at the $g_1=g_2$ unification scale $\Lambda_U$, 
and are
unambiguously measurable given knowledge of the superpartner spectrum
at the low scale and of course
the beta functions of the MSSM up to the $\Lambda_U$ scale.  Uncertainties
in the extracted $\delta$'s from measurements would spring from uncertainties
in superpartner masses and couplings, uncertainties in the all-orders
beta functions in the renormalization group evolution, and uncertainties
in the measured gauge couplings, most especially $\alpha_s$.

The values of $\delta_{1-2}$ and $\delta_{3-2}$ extracted from measurement
will have discriminating power in the GUT scale parameter space of
minimal supersymmetric $SU(5)$.  In this sense, we are  testing
the theory.  There are four parameters of the GUT theory that are
affecting the ratios of the
gaugino mass values at $\Lambda_U$,
\beq
\epsilon,~~ {A_\Sigma}/{\mbar},~~ {B_\Sigma}/{\mbar},~~ {B_5}/{\mbar}.
\eeq
Fitting four parameters to the two $\delta$ observables does not sound
particularly enlightening, but there are a few interesting observations
one can make about the underlying GUT model and the $\delta$ values.

For example, in minimal $SU(5)$ there is a relationship between $A$ terms
and $B$ terms that must be satisfied in order to solve the doublet-triplet
splitting problem,
\beq
A_\Sigma - B_\Sigma =A_\lambda -B_5 .
\label{AB terms}
\eeq
One solution to realize this relationship is the hypothesis of
universal $A$-terms $(A_\Sigma=A_\lambda\equiv A)$ and $B$-terms
$(B_\Sigma =B_5\equiv B)$. Under this hypothesis, possible regions 
of $\delta_{1-2}$ and $\delta_{3-2}$ are shown in Fig.~\ref{d12_d32}
with $\epsilon=0,3,5$ and $10\%$ assuming $|A/\overline{M}|<3$
and $|B/\overline{M}|<3$.
As one can see from Fig.~\ref{d12_d32}, 
a relative sign between
$\delta_{1-2}$ and $\delta_{3-2}$ tends toward negative in the
$\epsilon=0$ case, and toward positive in the non-zero $\epsilon$
cases. Also the $\delta$ corrections can be larger as $\epsilon$
gets larger. Therefore, there is an interesting opportunity
to unveil a crucial role of the non-zero $\epsilon$ effect
if we achieve precise enough determinations of gaugino masses at $\Lambda_U$.

The $\epsilon$ effect on gaugino masses is an important one.  Without
it, the corrections to the gaugino mass ratios at the high scale fall along
the rather narrow $\epsilon=0\%$ band in Fig.~\ref{d12_d32}.
Non-zero $\epsilon$ means, for example, that both $\delta_{1-2}$
and $\delta_{3-2}$ can be large and negative which is otherwise impossible.

There are two interesting limits
to consider to illustrate how patterns of fundamental parameters
can alter expectations of gaugino masses.  One limit is when $A=B\simeq 0$ 
and the
only corrections to the gaugino masses come from $\overline{M}$ corrections.
In that case, both $\delta_{1-2}$ and $\delta_{3-2}$ are approximately
$-1\%$, a negative but small mismatch of gaugino masses at the high scale.
Measuring these parameters to the sub-percent level is challenging
even for a linear collider.  We will outline measurement prospects
below. In any event, it would perhaps be easier
to rule out $\delta_{1-2}\simeq\delta_{3-2}\simeq-1\%$ than confirm
it by experiment.  Thus, if both $\delta$'s are positive or one
has a magnitude
much bigger than $1\%$, we will know that minimal $SU(5)$
with negligible $A$ and $B$ terms is not supported by the data.

The other interesting limit that eliminates the $\epsilon$ effect
on the gaugino mass ratios is when $A-B\simeq 0$, but $A$ and $B$ are
non-zero. Since the $\epsilon$ contribution always is prefactored
by $A-B$, the $\epsilon$ value has no effect on the gaugino mass ratios
in this case.  Thus, variations of
$A (=B)$ over its full range yields a line going through the origin
that connects the two
multi-line intersections in Fig.~\ref{d12_d32}.  That range
is characterized by
\beq
-4\% \lsim \delta_{1-2}\lsim 2\%~~~{\rm and}~~~
-6\% \lsim \delta_{2-3}\lsim 3\% .
\eeq
Therefore, any deviations beyond 10\% would be firm evidence against this
scenario, and even ${\cal O}(1\%)$ effects
that deviate from the $A=B\neq 0$ line would disaffirm the hypothesis.

Finally, we comment on the prospects of measuring $\delta_{1-2}$
and $\delta_{3-2}$ to the precisions required to make interesting 
suppositions about minimal $SU(5)$.  Very precise measurements of
all superpartner masses and couplings are crucial.  
Given precise measurements of these quantities at the low scale,
the
scale $\Lambda_U$ can be derived unambiguously.  The two-loop
evaluation of the gaugino
masses up to this scale is a well-defined 
prescription~\cite{two loop}.
Blair {\em et al.}~\cite{Blair:2000gy} have shown that a high-energy
linear collider is capable of measuring gaugino masses well enough
at the low-scale that a $\delta_{1-2}$ measurements at even the percent level
can be discerned. Measuring
$\delta_{3-2}$ down to this accuracy is not as easy, but it appears possible
that even $\delta_{3-2}\sim $ few percent could be established given careful
analysis of LHC and linear collider data.  The studies of 
Ref.~\cite{Blair:2000gy} are very encouraging in that we believe they
show that a linear collider along with the LHC could
make a significant impact on our ability to draw interesting distinctions
between GUT scale theories.

In conclusion, we have seen that no analysis of GUT gauge coupling
unification can be complete without taking into account NRO corrections
to the gauge kinetic function, and the expected size of these
corrections from naive dimensional analysis suggests that they
can play a decisive role in whether or not a theory is even viable.
This is the case for minimal supersymmetric $SU(5)$, where the
adjoint-Higgs NRO corrections can save the theory.  Furthermore, we
have shown that there are experimental consequences at the low scale,
and have illustrated how
careful measurements of the gaugino mass spectrum can discern ideas,
such as whether minimal $SU(5)$ with 
the universal $A$-term and $B$-term is viable.
Further theoretical and experimental ideas would
then be required to more definitively establish the theory or falsify it.

\section*{Acknowledgements}
This work was supported in part by the Department of Energy and the
Alfred P. Sloan Foundation.


\begin{thebibliography}{99}

\bibitem{Sakai:1981pk}
N.~Sakai and T.~Yanagida,
``Proton Decay In A Class Of Supersymmetric Grand Unified Models,''
Nucl.\ Phys.\ B {\bf 197}, 533 (1982);
S.~Weinberg,
``Supersymmetry At Ordinary Energies. 1. Masses And Conservation Laws,''
Phys.\ Rev.\ D {\bf 26}, 287 (1982).

\bibitem{Goto:1998qg}
T.~Goto and T.~Nihei,
``Effect of RRRR dimension 5 operator on the proton decay in the minimal
SU(5) SUGRA GUT model,'' 
Phys.\ Rev.\ D {\bf 59}, 115009 (1999)
[hep-ph/9808255].

\bibitem{Murayama:2001ur}
H.~Murayama and A.~Pierce,
``Not even decoupling can save minimal supersymmetric SU(5),''
Phys.\ Rev.\ D {\bf 65}, 055009 (2002)
[hep-ph/0108104].

\bibitem{Raby:2002wc}
S.~Raby,
``Proton decay,''
hep-ph/0211024 and references therein.



\bibitem{Bajc:2002bv}
B.~Bajc, P.~F.~Perez and G.~Senjanovic,
``Proton decay in minimal supersymmetric SU(5),''
Phys.\ Rev.\ D {\bf 66}, 075005 (2002)
[hep-ph/0204311];
B.~Bajc, A.~Melfo and G.~Senjanovic,
``Proton decay and fermion masses in supersymmetric grand unified  theories,''
hep-ph/0304051.



\bibitem{Hisano:1994fn}
J.~Hisano, T.~Moroi, K.~Tobe and T.~Yanagida,
``Suppression of proton decay in the missing partner model for
supersymmetric SU(5) GUT,'' 
Phys.\ Lett.\ B {\bf 342}, 138 (1995)
[hep-ph/9406417];
Y.~Kawamura,
``Triplet-doublet splitting, proton stability and extra dimension,''
Prog.\ Theor.\ Phys.\  {\bf 105}, 999 (2001)
[hep-ph/0012125];
L.~J.~Hall and Y.~Nomura,
``Gauge unification in higher dimensions,''
Phys.\ Rev.\ D {\bf 64}, 055003 (2001)
[hep-ph/0103125];
G.~Altarelli and F.~Feruglio,
``SU(5) grand unification in extra dimensions and proton decay,''
Phys.\ Lett.\ B {\bf 511}, 257 (2001)
[hep-ph/0102301].

\bibitem{Hill:1983xh}
C.~T.~Hill,
``Are There Significant Gravitational Corrections To The Unification Scale?''
Phys.\ Lett.\ B {\bf 135}, 47 (1984).

\bibitem{Shafi:gz}
Q.~Shafi and C.~Wetterich,
``Modification Of GUT Predictions In The Presence Of Spontaneous Compactification,''
Phys.\ Rev.\ Lett.\  {\bf 52}, 875 (1984).

\bibitem{Drees:1985bx}
M.~Drees,
``Phenomenological Consequences Of N=1 Supergravity Theories With Nonminimal Kinetic Energy Terms For Vector Superfields,''
Phys.\ Lett.\ B {\bf 158}, 409 (1985);
M.~Drees,
``N=1 Supergravity Guts With Noncanonical Kinetic Energy Terms,''
Phys.\ Rev.\ D {\bf 33}, 1468 (1986).

\bibitem{Ellis:1985jn}
J.~R.~Ellis, K.~Enqvist, D.~V.~Nanopoulos and K.~Tamvakis,
``Gaugino Masses And Grand Unification,''
Phys.\ Lett.\ B {\bf 155}, 381 (1985).

\bibitem{Hall:1992kq}
L.~J.~Hall and U.~Sarid,
``Gravitational smearing of minimal supersymmetric unification predictions,''
Phys.\ Rev.\ Lett.\  {\bf 70}, 2673 (1993)
[hep-ph/9210240].

\bibitem{Anderson:1996bg}
G.~Anderson {\em et al.},
``Motivations for and implications of non-universal GUT-scale boundary  conditions for soft SUSY-breaking parameters,''
hep-ph/9609457.

\bibitem{Huitu:1999eh}
K.~Huitu, Y.~Kawamura, T.~Kobayashi and K.~Puolamaki,
``Generic gravitational corrections to gauge couplings in SUSY SU(5)  GUTs,''
Phys.\ Lett.\ B {\bf 468}, 111 (1999)
[hep-ph/9909227].


\bibitem{Emmanuel-Costa:2003pu}
D.~Emmanuel-Costa and S.~Wiesenfeldt,
``Proton decay in a consistent supersymmetric SU(5) GUT model,''
Nucl.\ Phys.\ B {\bf 661}, 62 (2003)
[hep-ph/0302272];
M.~Kakizaki and M.~Yamaguchi,
``U(1) flavor symmetry and proton decay in supersymmetric standard model,''
JHEP {\bf 0206}, 032 (2002)
[hep-ph/0203192].





\bibitem{Bagger:1995bw}
See, for example, J.~Bagger, K.~T.~Matchev and D.~Pierce,
``Precision corrections to supersymmetric unification,''
Phys.\ Lett.\ B {\bf 348}, 443 (1995)
[hep-ph/9501277].

\bibitem{Hisano:1992mh}
J.~Hisano, H.~Murayama and T.~Yanagida,
``Probing GUT scale mass spectrum through precision measurements on the
weak scale parameters,'' 
Phys.\ Rev.\ Lett.\  {\bf 69}, 1014 (1992);
``Nucleon decay in the minimal supersymmetric SU(5) grand unification,''
Nucl.\ Phys.\ B {\bf 402}, 46 (1993)
[hep-ph/9207279].

\bibitem{Hisano:1993zu}
See, for example,
J.~Hisano, H.~Murayama and T.~Goto,
``Threshold correction on gaugino masses at grand unification scale,''
Phys.\ Rev.\ D {\bf 49}, 1446 (1994).

\bibitem{two loop}
S.~P.~Martin and M.~T.~Vaughn,
``Two loop renormalization group equations for soft supersymmetry breaking couplings,''
Phys.\ Rev.\ D {\bf 50}, 2282 (1994)
[hep-ph/9311340];
Y.~Yamada,
``Two loop renormalization group equations for soft SUSY breaking scalar interactions: Supergraph method,''
Phys.\ Rev.\ D {\bf 50}, 3537 (1994)
[hep-ph/9401241];
I.~Jack and D.~R.~Jones,
``Soft supersymmetry breaking and finiteness,''
Phys.\ Lett.\ B {\bf 333}, 372 (1994)
[hep-ph/9405233].


\bibitem{Blair:2000gy}
G.~A.~Blair, W.~Porod and P.~M.~Zerwas,
``Reconstructing supersymmetric theories at high energy scales,''
Phys.\ Rev.\ D {\bf 63}, 017703 (2001)
[hep-ph/0007107];
G.~A.~Blair, W.~Porod and P.~M.~Zerwas,
``The reconstruction of supersymmetric theories at high energy scales,''
Eur.\ Phys.\ J.\ C {\bf 27}, 263 (2003)
[hep-ph/0210058].

\end{thebibliography}
\end{document}